\documentclass[aps,prd,showkeys,nofootinbib,amsmath,amssymb,superscriptaddress,onecolumn]{revtex4-2}
\usepackage{dsfont}
\usepackage{mathtools} 
\usepackage{xcolor}
\definecolor{nicegreen}{rgb}{0.07, 0.564, 0.04}
\usepackage{framed}
\usepackage{comment}
\usepackage{xspace}
\usepackage{graphicx}
\usepackage{dcolumn}
\usepackage{bm}
\usepackage{tikz-cd}
\usepackage[normalem]{ulem}

\newcommand\ie{\mbox{\textit{i.\,e.}}\xspace}

\newcommand{\hx}{\hat{x}}
\newcommand{\hp}{\hat{p}}
\newcommand{\hpi}{\hat{\pi}}
\newcommand{\hX}{\hat{X}}

\newcommand{\hPi}{\hat{\Pi}}
\newcommand{\id}{\mathds{1}}
\newcommand{\Ord}{\mathcal{O}}

\DeclarePairedDelimiter\braket{\langle}{\rangle}
\DeclarePairedDelimiterX\Braket[2]{\langle}{\rangle}{#1 \delimsize\vert #2}

\begin{document}




\title{Spin operator, Bell nonlocality and Tsirelson bound in quantum-gravity induced minimal-length quantum mechanics}

\author{Pasquale Bosso} 
\affiliation{University of Lethbridge, Department of Physics and Astronomy, 4401 University Drive, Lethbridge, T1K 3M4, Alberta, Canada}
\affiliation{Dipartimento di Ingegneria Industriale, Universit\`a degli Studi di Salerno, Via Giovanni Paolo II, 132 I-84084 Fisciano (SA), Italy}
\author{Luciano Petruzziello}
\affiliation{Dipartimento di Ingegneria Industriale, Universit\`a degli Studi di Salerno, Via Giovanni Paolo II, 132 I-84084 Fisciano (SA), Italy}
\affiliation{INFN, Sezione di Napoli, Gruppo collegato di Salerno, Via Giovanni Paolo II, 132 I-84084 Fisciano (SA), Italy}
\affiliation{Institut f\"ur Theoretische Physik, Albert-Einstein-Allee 11, Universit\"at Ulm, 89069 Ulm, Germany}
\author{Fabian Wagner}
\affiliation{Institute of Physics, University of Szczecin, Wielkopolska 15, 70-451 Szczecin, Poland}
\affiliation{Dipartimento di Ingegneria Industriale, Universit\`a degli Studi di Salerno, Via Giovanni Paolo II, 132 I-84084 Fisciano (SA), Italy}
\author{Fabrizio Illuminati}
\email[Corresponding author: ]{filluminati@unisa.it}
\affiliation{Dipartimento di Ingegneria Industriale, Universit\`a degli Studi di Salerno, Via Giovanni Paolo II, 132 I-84084 Fisciano (SA), Italy}
\affiliation{INFN, Sezione di Napoli, Gruppo collegato di Salerno, Via Giovanni Paolo II, 132 I-84084 Fisciano (SA), Italy}

\date{April 26, 2023}

\begin{abstract}
\begin{center}
    \bf Abstract
\end{center}
Different approaches to quantum gravity converge in predicting the existence of a minimal scale of length. This raises the fundamental question as to whether and how an intrinsic limit to spatial resolution can affect quantum mechanical observables associated to internal degrees of freedom. We answer this question in general terms by showing that the spin operator acquires a momentum-dependent contribution in quantum mechanics equipped with a minimal length. Among other consequences, this modification induces a form of quantum nonlocality stronger than the one arising in ordinary quantum mechanics. In particular, we show that violations of the Bell inequality can exceed the maximum value allowed in ordinary quantum mechanics, the so-called Tsirelson bound, by a positive-valued function of the momentum operator. We introduce possible experimental settings based on neutron interferometry and quantum contextuality, and we provide preliminary estimates on the values of the physical parameters needed for actual laboratory implementations.   
\end{abstract}

\maketitle



\noindent
\section{Introduction}
The development of a coherent and self--consistent theory of quantum gravity appears as one of the most ambitious and controversial research goals in contemporary physics.
Despite the existence of promising candidate theories, progress in their development and assessment is hampered by the difficulty of probing genuine signatures of quantum gravitational effects at the scales of energy, length and time required in laboratory tests or in astrophysical observations.

In this context, much effort has been devoted to the construction of physically motivated effective models with the aim of identifying, quantifying and ideally testing specific features expected to emerge in the regime where quantum and gravitational effects can be simultaneously relevant \cite{Amelino-Camelia:2008aez}.
A significant number of such models imply the existence of a minimal measurable length, thus forbidding the possibility of localizing quantum systems with arbitrary precision \cite{Gross:1987kza,Gross:1987ar,Amati:1987wq,Amati:1988tn,Konishi:1989wk,Garay:1994en,Adler:1999bu,Scardigli:1999jh,Capozziello:1999wx,Padmanabhan:2015vma}. 
On the other hand, a theme of central interest that has stimulated an intense and ongoing debate concerns the relations between quantum gravity and the defining properties of quantum entanglement and quantum nonlocality, since specific aspects of the latter might provide indirect evidence of the former \cite{sorkin,popescu,dowker,erepr,bose,vedral}. 

Given this framework, we analyze how the existence of a minimal length affects the structure of quantum mechanical observables and nonlocal quantum correlations when moving from the high-energy regimes to non-relativistic quantum mechanics. 
In so doing, we show that in  minimal-length quantum mechanics (MLQM) the internal degrees of freedom are modified by acquiring a functional dependence on the motional ones. In turn, this generalization significantly impacts quantum nonlocality and Bell's theorem of ordinary quantum mechanics (QM) \cite{Bell1964,Bell2004,Brunner2014}. 
In particular, we find that the Tsirelson upper bound to the violation of Bell inequalities in ordinary QM \cite{Cirelson:1980ry,Tsirelson:1987,axiom1,axiom2,axiom3} is enhanced in MLQM by a positive multiplicative function of the motional degrees of freedom, thereby yielding a form of quantum nonlocality stronger than in ordinary QM.

In principle, such a quantitative prediction is experimentally testable and could be exploited to prove or disprove the existence of a minimal scale of length. Moreover, since the generalized quantum mechanical observables and quantum nonlocal correlations depend on the model of choice, they give rise to different modifications of the Tsirelson bound, thus potentially allowing for the comparison and discrimination of distinct approaches to quantum gravity.

In order to provide some first rudimentary estimates on the possible experimental verification of these effects, we have supplemented the theoretical discussion with the description of a table-top laboratory test involving neutron interferometry \cite{rauch} and quantum contextuality \cite{context}. According to the latter, for any realistic hidden variable theory, all observables associated to a given system (belonging to both commuting and non-commuting sets) have preexisting, definite values that do not depend on the choice of the quantities being measured. Therefore, this notion heavily relies on the assumption of realism already discussed in the original formulation of Bell's theorem \cite{Bell1964}; on the other hand, it does not require the locality axiom, thus making Bell nonlocality a special case in the broader arena of quantum contextuality. By virtue of this generality,
one can devise an experiment measuring the correlations between the distinct degrees of freedom of a single system (in our case, the neutron), thus avoiding the much more laborious preparation of a multipartite initial state \cite{inter}.

\section{Results}

\subsection{Deformed commutation relations and generalized uncertainty principle}

Assuming the existence of a minimal length, in the non-relativistic limit several phenomenological models of quantum gravity reduce to ordinary quantum mechanics with deformed canonical commutation relations (DCCRs) and, correspondingly, a generalized uncertainty principle (GUP). 

A physically and mathematically consistent starting point is the Robertson--Schr\"odinger prescription for non-commuting observables \cite{Robertson29,Schroedinger30}, which allows to infer modifications of the Heisenberg algebra from an uncertainty relation featuring a minimal scale of length \cite{Maggiore:1993rv,Kempf:1994su,Das:2008kaa,Ali:2011fa,nouicer,Pedram:2011gw,Husain:2012im,Bosso:2017ndq,Shababi:2017zrt,Bosso:2020aqm,Wagner:2021bqz,Petruzziello21,natur,natur2,pedram3,shababi2,fadel}.
In particular, concerning the most general scheme that includes coordinate commutativity, isotropy and rotational invariance, it can be shown that the \emph{physical} phase-space variables satisfy the algebra \cite{Maggiore:1993kv,Kempf:1994su}
\begin{subequations}\label{GUP}
    \begin{align}
        \left[\hx^i,\hx^j\right]=&\left[\hpi_i,\hpi_j\right]=0 \, , \label{eqn:commutativity}\\
        \left[\hx^i,\hpi_j\right] =& \, i\left[f\left(\hpi^2\right)\delta^i_j + g\left(\hpi^2\right)\frac{\hpi^i\hpi_j}{\hpi^2}\right] \, ,
    \end{align}
\end{subequations}
with $f$ and $g$ being two arbitrary, analytic, dimensionless functions and where $\hpi^2\equiv\delta^{ij}\hpi_i\hpi_j$ (here and in the following, unless otherwise stated, the use of natural units is understood throughout).
Verification of the only nontrivial Jacobi identity 
$$\left[\hat{x}^i,\left[\hat{x}^j,\hat{\pi}_k\right]\right]+\mathrm{cycl.\,perm.}=0$$
constrains the above functions to satisfy the relation
\begin{equation}
    g = \frac{2\left(\log f\right)'\hpi^2}{1-2\left(\log f\right)'\hpi^2}f \, ,
    \label{noncomcond}
\end{equation}
where the prime denotes derivation with respect to $\hpi^2$. This condition is consistent with the requirement of spatial commutativity only if
\begin{equation}
    2(\log f)'\hpi^2<1, \qquad \forall \, \, \, \hpi^2 \, . \label{noncomcond2}
\end{equation} 
A comment is in order here: coordinate non-commutativity and minimal length are in principle two independent aspects.
Distinct models may feature minimal length and no coordinate non-commutativity, no minimal length and coordinate non-commutativity, or both minimal length and coordinate non-commutativity at the same time. The first and the third type of models can be directly implemented by either fulfilling or violating the identity \eqref{noncomcond}, respectively. To mention a simple example of the second kind, one can consider the set of commutators $[\hx^i,\hx^j] = i \theta^{ij}$ with $\theta^{ij}$ constants and $[\hx^i,\hp_j] = i \delta^i_j$. Such a scheme evidently does not predict any minimal length, but does encompass spatial non-commutativity. 

Now, in order to define proper observables associated to the internal degrees of freedom (such as the spin) and the corresponding notion of nonlocality in a MLQM equipped with DCCRs and a GUP, we need to verify that the classical limit of such extended quantum mechanics coincides with the one of ordinary quantum mechanics, so as to guarantee the possibility of local hidden variable theories and the validity of Bell inequalities in both settings.
The classical limit of effective schemes encompassing DCCRs and a GUP leads to standard, unmodified Poisson brackets at the perturbative level of small corrections to the canonical commutation relations \cite{Casadio:2020rsj}.
Here, the same result applies to the general case of corrections behaving as a function $f(\hpi^2)$ of the momentum operator squared.

Since the functions $f$ and $g$ must be dimensionless, their argument must be made dimensionless by introducing a characteristic energy scale. Such quantum gravitational threshold is commonly identified by the Planck mass $m_\text{p}=\sqrt{\hbar \,c/G_{\text{N}}}$, with $\hbar$ being the reduced Planck constant, $G_N$ the gravitational constant and $c$ the speed of light in vacuum, so that the argument of the function $f$ is rescaled accordingly
\begin{equation}\label{dep}
f = f\left(\frac{\hpi^2}{m_\text{p}^2c^2}\right) \, .
\end{equation}
Ordinary QM is recovered in the low-energy limit, yielding $\lim_{\hpi^2\rightarrow 0}f=1$ and $\lim_{\hpi^2\rightarrow 0} g = 0$. 
Due to the functional dependence of Eq. \eqref{dep}, the above limit is equivalent to the formal one $G_N\rightarrow 0$, which amounts to neglect the presence of gravity.
Similarly, a classical dynamical theory (CT) is retrieved in the formal limit $\hbar\rightarrow 0$.
Since in Eq. \eqref{dep} $G_N$ comes in pair with the inverse of $\hbar$, the resulting CT is independent of the gravitational constant. Indeed, for $G_N\rightarrow 0$ and $\hbar\rightarrow 0$ simultaneously, the Poisson brackets stemming from the DCCRs can only either be divergent/vanishing (two ill-defined scenarios) or exhibit a constant correction. In the latter case, they are related to the Poisson brackets of classical mechanics by a canonical transformation; correspondingly, a well-defined classical limit is necessarily trivial.

The above discussion may be summarized by the following illustrative diagram:
\begin{equation}
\begin{tikzcd}
\text{MLQM} \arrow[rd,"\hbar\rightarrow 0"'] \arrow[r, "G_{\text{N}}\rightarrow 0"] & \text{QM} \arrow[d,"\hbar\rightarrow 0"] \\
& \text{CT}
\end{tikzcd}\label{limitdiag}
\end{equation}
{which underpins the non-commutativity of the limits $G_{\text{N}}\rightarrow 0$ and $\hbar\rightarrow 0$.} 

After having established the well-defined behavior of the classical limit, we move to discuss how Eqs. \eqref{GUP} entail the presence of a minimal scale of length. Recalling the Robertson--Schr\"odinger derivation of the uncertainty relation for two non-commuting observables $\hx^i$ and $\hpi_i$, namely
\begin{equation}
    \Delta x^i
    \geq \frac{|\langle[\hx^i,\hpi_i]\rangle|}{2(\Delta \pi_i)}
    = \frac{1}{2\Delta \pi_i} \left|\langle f(\hpi^2) \rangle + \left\langle g(\hpi^2) \frac{(\hpi^i)^2}{\hpi^2}\right\rangle\right| \, ,
    \label{eqn:RS}
\end{equation}
we note that, depending on the choice of $f$, the r.h.s. of Eq. \eqref{eqn:RS} may feature a global minimum, thus implying a nonvanishing minimal length. We obtain an immediate visualization of this occurrence by expanding $f$ and $g$ to leading order in $\hat{\pi}^2/m_\text{p}^2$ in a generic scheme of MLQM
\begin{align}
    f(\hpi^2) \simeq &\, 1 + \beta \frac{\hpi^2}{m_\text{p}^2} \, , &
    g(\hpi^2) \simeq &\, 2 \beta \frac{\hpi^2}{m_\text{p}^2} \, ,\label{quadraticmodel}
\end{align}
where the dimensionless deformation parameter $\beta$ characterizes the specific model considered.
Then, for the relevant case $\beta >0$, to leading order:
\begin{equation}
    \left|\langle f(\hpi^2) \rangle + \left\langle g(\hpi^2) \frac{(\hpi^i)^2}{\hpi^2}\right\rangle\right|
    \gtrsim \left|1 + \beta \frac{\sum_j\Delta \pi^j\Delta \pi_j}{m_\text{p}^2} + 2 \beta \frac{(\Delta \pi^i)^2}{m_\text{p}^2}\right| \, ,
\end{equation}
but since $\sum_j\Delta\pi_j\Delta\pi^j\geq (\Delta\pi_i)^2$, it follows that
\begin{equation}
    \Delta x^i \geq \frac{1}{2 \Delta\pi_i}\left[1+3\beta \left(\frac{\Delta\pi_i}{m_{\text{p}}}\right)^2\right] \, .
\end{equation}
This expression features a global minimum for \mbox{$\Delta x^i = \sqrt{3 \beta}/m_{\text{p}}$}, which encompasses the existence of a limited spatial resolution.

Returning to Eqs. \eqref{GUP} and resorting to the Einstein convention on repeated indices, we may introduce suitable auxiliary operators in order to recover the canonical symplectic structure. We thus define the \emph{canonical} variables $\hat{X}^i$ and $\hat{\Pi}^i$ as
\begin{equation}
    \hX^i=\hx^i,\hspace{1cm}\hPi_i=\frac{\hpi_i}{f} \, ,
    \label{eqn:auxiliary}
\end{equation}
which satisfy the Heisenberg algebra
\begin{equation}
\left[\hX^i,\hX^j\right]=\left[\hPi_i,\hPi_j\right]=0 \, ,\hspace{1cm}\left[\hX^i,\hPi_j\right]=i\delta^i_j \, .
\end{equation}
%
The construction of canonical and physical operators extends to angular momentum \cite{angularmomentum}. In the context of a modified quantum mechanics with DCCRs, the \emph{physical} orbital angular momentum operator is defined as $\hat{l}_i\equiv\epsilon_{ijk}\hat{x}^j\hat{\pi}^k$. This operator satisfies the deformed algebra 
\begin{equation}\label{defalg}
\left[\hat{l}_i,\hat{l}_j\right]=i\varepsilon_{ijk}\hat{l}^kf\left(\hat{\pi}^2\right) \, .
\end{equation}
The ordinary SO(3) algebra is recovered by introducing the \emph{canonical} angular momentum operator 
\begin{equation}\label{aux}
\hat{L}_i=\frac{\hat{l}_i}{f} \, .
\end{equation}
Note that, since $\hat{\pi}^2$ is the Casimir invariant of the algebra, there is no operator ordering ambiguity. These considerations suggest an analogous reformulation of the intrinsic angular momentum in MLQM.

\subsection{Spin operator in the presence of a minimal scale of length}

Let us consider a spinor field interacting with a classical magnetic field. In momentum space, the Dirac equation describing a relativistic spin-$1/2$ particle reads 
\begin{equation}
    i\partial_t\psi=\left(\gamma^i\hpi_i+\gamma^0m\right)\psi \, ,
    \label{eqn:Dirac_physical}
\end{equation}
with the gamma matrices in the Dirac representation
\begin{equation}
\gamma^i=\begin{pmatrix}
0&\sigma^i\\
\sigma^i&0
\end{pmatrix},\hspace{1cm}\gamma^0=\begin{pmatrix}
\id&0\\
0&-\id
\end{pmatrix}.
\end{equation}
In order to account for an external magnetic field such that $A_\mu=\left(0,A_i\right)$ \cite{drell}
and by means of the minimal coupling prescription, the physical momenta in Eq. \eqref{eqn:Dirac_physical} are replaced by the operators $\hp_i=\hpi_i-eA_i$ that preserve the $U(1)$ gauge invariance.
This transformation modifies the underlying gauge symmetry in such a way that the theory remains invariant under the modified gauge transformations. 

Splitting the spinor field into its particle/antiparticle components $\psi=(\varphi,\chi)$, we can consider the non-relativistic limit by singling out a mass-dependent phase $\psi\rightarrow e^{-imt}\psi$ and employing the non-relativistic limit $|m\psi|\gg |\partial_t\psi|.$
Within this framework, the generalized Schr\"odinger equation governing the dynamics of the particle wave function $\varphi$ in MLQM reads
\begin{equation}
i\partial_t\varphi=\frac{\sigma^i\hp_i\sigma^j\hp_j}{2m}\varphi \, .
\end{equation}
Upon introducing the magnetic field $B^i=\epsilon^{ijk}\partial_jA_k$ along with the spin operator $\hat{S}_i=\sigma_i/2,$ we obtain that, up to $\mathcal{O}(e)$, the effective state vector dynamics in MLQM reads
\begin{equation}\label{schro}
    i\partial_t\varphi 
    = \left[\frac{f^2}{2m}\hat{\Pi}^2 - ef\left(\hat{L}_i+2\hat{S}_i\right)B^i\right] \varphi 
    = \left[\frac{\hat{\pi}^2}{2m} - e\left(\hat{l}_i+2\hat{s}_i\right)B^i\right] \varphi \, ,
\end{equation}
where we have defined the deformed spin operator as $\hat{s}_i\equiv f \hat{S}_i.$ It is worth observing that, when written in terms of the physical operators, no correction proportional to $f$ appears in the dynamical equation.


In light of the above, the physical intrinsic and orbital angular momentum operators that the external magnetic field couples to are, respectively, $\hat{s}_i$ and $\hat{l}_i$. Consequently, the deformation of the spin algebra coincides with the deformation of the orbital angular momentum algebra \eqref{defalg}, namely
\begin{equation}\label{spina}
\left[\hat{s}_i,\hat{s}_j\right]=i\hbar\varepsilon_{ijk}\hat{s}^kf\left(\hat{\pi}^2\right) \, .
\end{equation}
This observation implies that expectation values of spin-dependent observables on any quantum state must account for the presence of the momentum operator.
Therefore, given a quantum state that is a tensor product of spin- and momentum-dependent components, \ie: $|\Psi\rangle=|\psi_s\rangle\otimes|\psi_\pi\rangle$, we have that in MLQM
\begin{equation}\label{action}
\hat{s}_i|\Psi\rangle=\hat{S}_i|\psi_s\rangle\otimes f(\hat{\pi}^2)|\psi_\pi\rangle \, ,
\end{equation}
which yields
\begin{equation}\label{ev}
\langle\hat{s_i}\rangle = \braket{f(\hat{\pi}^2)}\langle\hat{S}_i\rangle \, .    
\end{equation}

\subsection{Quantum nonlocality and Tsirelson bound} 

A most fundamental question arising in MLQM concerns how momentum-dependent state expectations of the form of Eq. \eqref{ev} modify quantum nonlocality as ruled by Bell's theorem \cite{Bell1964,Bell2004,Brunner2014}.

In the standard framework of Bell nonlocality, two parties, $A$ and $B$, perform causally separated experiments on a bipartite system.
Let $\hat{A}$ and $\hat{B}$ denote two dichotomous observables associated with party $A$ and party $B$, $(a,a')$ and $(b,b')$ two sets of different experimental settings associated with party $A$ and party $B$, respectively, and $C( \cdot , \cdot )$ the correlation functions between outcomes corresponding to the four available experimental configurations.
Within a local hidden variable theory, a suitable combination $\mathcal{S}$ of such correlations must be bounded according to the following Clauser-Horne-Shimony-Holt (CHSH) inequality \cite{Clauser:1969ny}:
\begin{equation}\label{chsh}
    \mathcal{S}=\left|C\left(a,b\right)-C\left(a,b'\right)+C\left(a',b\right)+C\left(a',b'\right)\right|\leq2 \, .  
\end{equation}
In the context of ordinary QM one has $C(a,b)=\langle\hat{A}(a)\hat{B}(b)\rangle$ (and analogously for the remaining correlations) and the inequality Eq. \eqref{chsh} is violated. Remarkably, although the highest possible, absolute algebraic value for $\mathcal{S}$ is $\mathcal{S}^{\text{max}} = 4$, within QM the maximum achievable value is $\mathcal{S}^{\text{max}}_{\text{QM}} = 2\sqrt{2}$, the so-called Tsirelson bound, as it
was proved for the first time by B. Tsirelson \cite{Cirelson:1980ry,Tsirelson:1987}.
This bound places a strong limit to the degree of nonlocality featured by standard QM.

Let us now consider an experimental realization involving two spin-$1/2$ observables. We can choose any one of the four maximally entangled, two-qubit Bell states  as the initial state, for instance $|\Psi\rangle=\left(|0,1\rangle-|1,0\rangle\right)/\sqrt{2}$. Next, consider the four dichotomous observables
\begin{align}
\hat{A}(a) = & \hat{S}_z\otimes\textbf{1} \, , &
\hat{A}(a') = & \hat{S}_x\otimes\textbf{1} \, , \label{obs}\\
\hat{B}(b) = & -\frac{\textbf{1}\otimes\left(\hat{S}_z + \hat{S}_x\right)}{\sqrt{2}} \, , &
\hat{B}(b') = & \frac{\textbf{1}\otimes\left(\hat{S}_z - \hat{S}_x\right)}{\sqrt{2}} \, . \label{obs2}
\end{align}
With this choice, one finds that $\mathcal{S}$ saturates exactly the Tsirelson bound $2\sqrt{2}$, thereby realizing the largest degree of nonlocal correlation achievable within standard QM and showing that maximal entanglement is equivalent to maximal Bell nonlocality on pure states. Various physical principles have been invoked to justify the existence of such a strict upper limit to nonlocal quantum mechanical correlations between spatially-separated events in ordinary QM \cite{axiom1,axiom2,axiom3}.

In the framework of MLQM, the basic structure of quantum nonlocality stands firm; in particular, Bell's theorem and the CHSH inequality of standard QM continue to hold unmodified by virtue of the non-commutativity of the classical and non-gravitational limits, as summarized in \eqref{limitdiag}. On the other hand, Tsirelson bound must be generalized, as the spin correlation functions in CHSH-type experiments become momentum-dependent.
Indeed, by virtue of Eq.~\eqref{ev} and upon replacing the canonical spin operators with the physical ones \mbox{$(\hat{S}_x,\hat{S}_z)\rightarrow(\hat{s}_x,\hat{s}_z)$} in Eqs. \eqref{obs} and \eqref{obs2}, one has \mbox{$C(a,b)=\braket{f^2(\hat{\pi}^2)}\langle\hat{A}(a)\hat{B}(b)\rangle$}, and similarly for the other correlations. Therefore, after some algebra, one finds:
\begin{equation}\label{tsir}
    \mathcal{S}_{\text{MLQM}}^{\text{max}} = \braket{f^2\left(\hat{\pi}^2\right)} \, \mathcal{S}^{\text{max}}_{\text{QM}} = 2\sqrt{2} \,\braket{f^2\left(\hat{\pi}^2\right)} \, .
\end{equation}
Equation \eqref{tsir} shows that non-relativistic MLQM derived from effective, low-energy models of quantum gravity features nonlocal quantum correlations that are stronger than the ones holding in standard, non-relativistic QM without minimal-length induced deformations.
We might speculate that the modified Tsirelson bound $\mathcal{S}_{\text{MLQM}}^{\text{max}}$ might saturate the absolute algebraic limit of $\mathcal{S}^{\text{max}} = 4$
as one approaches the Planck regime; however, the question must remain open, since the validity of the existing low-energy effective models of quantum gravity breaks down at the Planck scale. Furthermore, it is likely that in the context of MLQM the very definition of the absolute algebraic limit might need to be generalized accordingly.

Given the above words of caution, we can provide some explicit (albeit very preliminary and extremely conservative) estimates of the correcting factor $f(\hat{\pi}^2)$ in Eq. \eqref{tsir} for two relevant classes of model that can be found in the literature. Let us first consider the linear correction deduced from models of the Doubly Special Relativity type \cite{ms1,ms2}:
\begin{equation}\label{doubla}
    \braket{f^2(\hat{\pi}^2)}\simeq 1+2\beta_\ell\frac{\braket{\sqrt{\hat{\pi}^2}}}{m_{\text{p}}}=1+\beta_\ell\frac{\sqrt{2ME_{\text{kin}}}}{m_{\text{p}}} \, ,
\end{equation}
where $M$ is the particle mass, $E_\text{kin}$ the kinetic energy, and $\beta_\ell$ the linear deformation parameter which, in a worst-case scenario, is typically considered to be of order unity, \ie, $\beta_\ell\sim\mathcal{O}(1)$. It is worth stressing that the above scheme predicts not only the existence of a minimal length, but also the presence of a maximal momentum. However, one can prove that the framework introduced so far can be extended to arbitrary deformations of standard quantum mechanics that account for minimal/maximal length and/or minimal/maximal momentum.

Now, assuming for instance a CHSH-type experiment performed on systems of muons of mass \mbox{$M \sim \Ord (10^{-1})$}~GeV and kinetic energy scales \mbox{$E_{\text{kin}}\sim \Ord (10^{-2})$}~GeV, well within the non-relativistic regime, we find for the relative correction ratio in orders of magnitude
\begin{equation}
    \Delta\mathcal{S} = \frac{\mathcal{S}_{\text{MLQM}}^{\text{max}} - \mathcal{S}^{\text{max}}_{\text{QM}}}{\mathcal{S}^{\text{max}}_{\text{QM}}} \sim \mathcal{O}(10^{-20}) \, .
\end{equation}
On the other hand, if we consider a string-induced MLQM, we have
\begin{equation}\label{stringa}
    \braket{f^2(\hpi^2)}\simeq 1+2\beta_q \frac{\braket{\hat{\pi}^2}}{m_{\text{p}}^2} \, .
\end{equation}
Assuming again a worst-case scenario for the deformation parameter, i.e. $\beta_q\sim \Ord (1),$ the relative correction is $\Delta\mathcal{S}\sim \Ord (10^{-40})$. 
Obviously, considering larger values of the deformation parameter and higher energy scales approaching relativistic regimes leads in all cases to significantly larger corrections. 

\subsection{Experiments: interferometry and contextuality}

The above crude numerical estimates can be put in perspective by considering realistic scenarios involving neutron interferometry \cite{rauch} and quantum contextuality \cite{context}. Taking the neutron as the physical system of choice realizes an optimal compromise in order to realize the most convenient experimental conditions for spin measurements with the accuracy required to test quantum gravity effects.
In fact, the phenomenology of quantum gravity suffers from an unfavorable feature when composite systems are taken into account \cite{soccer,soccer2}. Indeed, minimal-length effects decrease with increasing number of constituents; this degradation typically scales with $1/N^\alpha$, with $\alpha = 2$ when quasi-rigid motion is assumed \cite{soccer2}, while various experimental bounds on its range of definition are also available \cite{kumar,kumar2}. On the other hand, using charged spin-$1/2$ elementary particles with $N=1$ poses significant challenges in the realization of Bell-like experiments; therefore neutrons, being made of only three constituents, represent the best choice also in view of the fact that neutron interferometry is a well-established and advanced field experimental physical technology.
Additionally, neutron interferometry has evolved into a standard method of experimentation within the phenomenology of quantum mechanics, especially quantum contextuality \cite{inter,inter2,inter3,inter4},
as the interferometric manipulation of coherent neutron beams readily allows for the creation of non-classical correlations between distinct degrees of freedom.
This trait turns out to be particularly useful, as it is significantly easier to prepare and control a single system in an initial state featuring entangled degrees of freedom rather than two or more initially entangled parties which are then spatially separated and analyzed independently, as in standard Bell-type experiments.

\begin{figure}[ht]
\centering
\includegraphics[width=\textwidth]{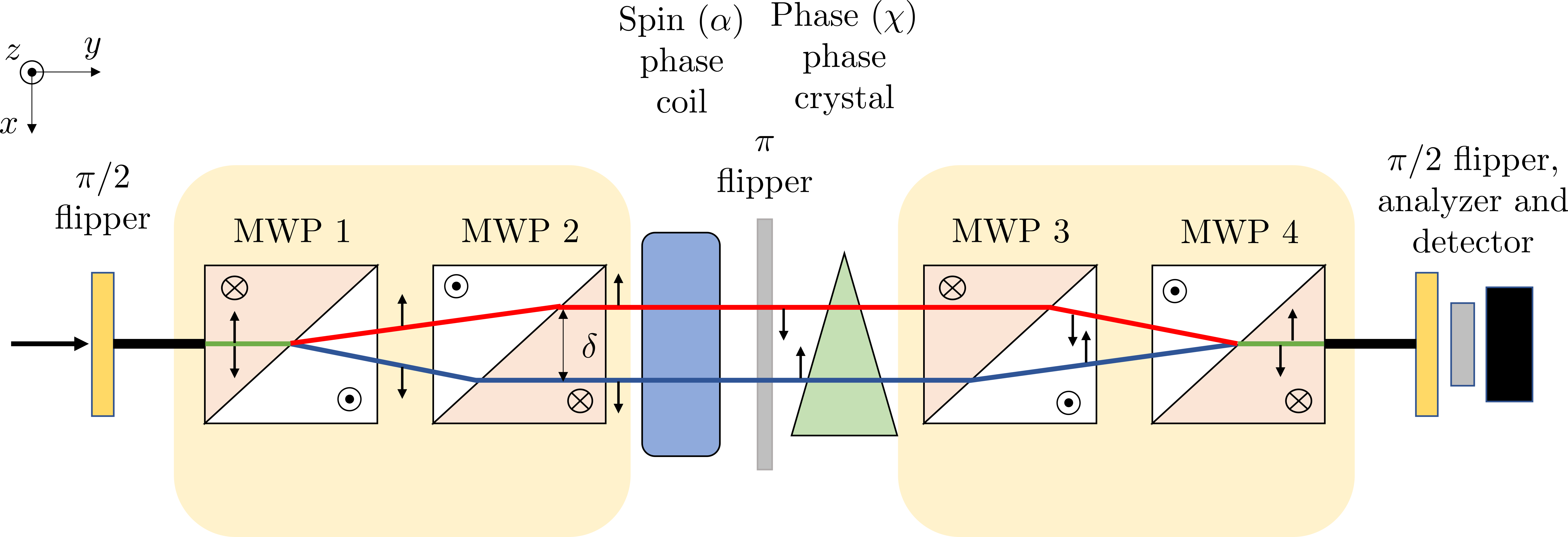}
\caption{Scheme of the proposed experimental setting. A beam of polarized neutrons passes through two MWPs to entangle spin states with path degrees of freedom. Before reverting the system to its initial condition with other two MWPs to allow for the detection process, the beam travels through a spin coil and a crystal to let the superposed states acquire a relative phase. Such phases modulate the different detector settings with which it is possible to define a CHSH-like inequality.}
\label{fig1}
\end{figure}

In light of the above, let us consider an experimental setting in which the internal and motional degrees of freedom of a neutron are initially correlated; to this aim, we consider the following laboratory setup \cite{inter2,inter3,inter4}, whose scheme is sketched in Fig. \ref{fig1}. A neutron beam is polarized in the $z$-direction while it propagates along the $y$-direction. By means of a $\pi/2$ flipper, the polarization of the beam is oriented along the $x$-direction. Then, the initial state can be viewed as a linear superposition in the spin $z$-basis
\begin{equation}
|\psi_{\mathrm{in}}\rangle=\frac{|\uparrow\rangle+|\downarrow\rangle}{\sqrt{2}}\otimes|g\rangle \, ,
\end{equation}
where $|g\rangle$ denotes the path state associated to the distance which is traveled by both $|\uparrow\rangle$ and $|\downarrow\rangle$ (see the green trajectory of Fig. \ref{fig1}).

\noindent
Subsequently, the neutron beam passes through two magnetic Wollaston prisms (MWPs), which are capable of separating the upper from the lower components of the spin state, allowing them to travel distinct paths separated by a definite distance $\delta,$ called entanglement length. The MWP is a cubic device divided into two triangular regions whose upper and lower faces are made up of superconducting coils. With these, it is possible to induce a strong magnetic field $\textbf{B}$ in each region. To achieve the splitting of $|\uparrow\rangle$ and $|\downarrow\rangle$, two antiparallel magnetic fields need to be created in the MWP, so as to give rise to a sharply localized field discontinuity at the interface. Such a discontinuity can be modeled by a quantum mechanical potential step, thereby explaining the bifurcation of the initial path.
After this process, the state $|\psi_{\mathrm{in}}\rangle$ has evolved into
\begin{equation}
|\psi_1\rangle=\frac{|\uparrow,r\rangle+|\downarrow,b\rangle}{\sqrt{2}}\,, 
\end{equation}
where $|\uparrow,r\rangle+|\downarrow,b\rangle\equiv|\uparrow\rangle\otimes|r\rangle+|\downarrow\rangle\otimes|b\rangle$, with $|r\rangle$ and $|b\rangle$ denoting the states associated with the red and the blue path, respectively (see Fig. \ref{fig1}). Next, by letting the neutron beam pass through a spin phase coil and a quartz crystal, a relative phase shift between the states of $|\psi_1\rangle$ can be introduced, namely
\begin{equation}
|\psi_2\rangle=\frac{|\uparrow,r\rangle+e^{i(\alpha+\chi)}|\downarrow,b\rangle}{\sqrt{2}}\,, 
\end{equation}
where $\alpha$ is the relative phase for the spin and $\chi$ the one for the path. Subsequently, the beam goes through another pair of MWPs, which this time acts as a disentangler, thereby modifying the state $|\psi_2\rangle$ into
\begin{equation}
|\psi_3\rangle=\frac{|\uparrow\rangle+e^{i(\alpha+\chi)}|\downarrow\rangle}{\sqrt{2}}\otimes|g\rangle\,.
\end{equation}
Finally, the neutrons come across another $\pi/2$ spin-turner before entering the polarization analyzer and the detector. One can then reconstruct the CHSH-like contextual witness $\mathcal{S'}$ from the number $N(\alpha,\chi)$ of neutrons detected for given phase shifts $\alpha$ and $\chi$. Indeed, in line with Eq. \eqref{chsh}, the quantity $\mathcal{S'}$ is written as
\begin{equation}\label{contesto}
\mathcal{S'}=\left|E(\alpha_1,\chi_1)+E(\alpha_1,\chi_2)+E(\alpha_2,\chi_1)-E(\alpha_2,\chi_2)\right|\,,    
\end{equation}
where the expectation value $E(\alpha,\chi)$ is defined as
\begin{equation}\label{bige}
E(\alpha,\chi)=\langle\psi_{\mathrm{in}}|\hat{\sigma}^s_\alpha\hat{\sigma}^p_\chi|\psi_{\mathrm{in}}\rangle\,, \qquad \hat{\sigma}^{s}_{\alpha}=2\left(\hat{S}_x\cos\alpha+\hat{S}_y\sin\alpha\right)\,, \qquad \hat{\sigma}^{p}_{\chi}=\hat{\sigma}^p_x\cos\chi+\hat{\sigma}^p_y\sin\chi\,.
\end{equation}
It is worth stressing that, whilst $\hat{S}_i$ are correctly identified with the spin observables since they are defined on the spin Hilbert space, the operators $\hat{\sigma}^p_i$ act on the path states, thus exhibiting no connection at all with spins.

The expectation values can now be recast in the following form \cite{inter}:
\begin{equation}\label{numb}
E(\alpha,\chi)=\frac{N(\alpha,\chi)-N(\alpha,\chi+\pi)-N(\alpha+\pi,\chi)+N(\alpha+\pi,\chi+\pi)}{N(\alpha,\chi)+N(\alpha,\chi+\pi)+N(\alpha+\pi,\chi)+N(\alpha+\pi,\chi+\pi)}\,.
\end{equation}
Bearing this in mind, we observe that, just like in the CHSH scenario, the inequality $\mathcal{S'}\leq2$ must hold in order to preserve a classical behavior. In fact, by selecting precisely $\alpha_1=0$, $\alpha_2=\pi/2$, $\chi_1=-\pi/4$ and $\chi_2=\pi/4$, one obtains $\mathcal{S'}=2\sqrt{2}$, 
\ie the maximal violation of the CHSH inequality that saturates the Tsirelson bound.

On the other hand, the numerical values obtained in concrete laboratory tests are slightly lower than the theoretically predicted ones.
This mismatch is prominently due to the overall dephasing/depolarization experienced by the neutron beam while passing through the spin flippers \cite{inter}. As a result, the maximum achievable violation turns out to be $\mathcal{S'}^{\mathrm{max}}_{\mathrm{QM}} \simeq 2.20$, while the classical behavior is lost as soon as $\mathcal{S'} > 1.56$ \cite{inter3}. Several experiments have achieved the largest possible value $\mathcal{S'}^{\text{max}}_{\text{QM}}$ with a precision of a part in a hundred \cite{inter,inter2,inter3,inter4}.

As shown in the previous section, upon replacing the canonical spins $\hat{S}_i$ with the physical ones $\hat{s}_i$, the quantum gravitational correction to the standard quantity $\mathcal{S'}^{\text{max}}_{\text{QM}}$ reads
\begin{equation}\label{expfin}
    \mathcal{S'}^{\text{max}}_{\text{MLQM}} = \langle f(\hat{\pi}^2)\rangle\mathcal{S'}^{\text{max}}_{\text{QM}} \, .
\end{equation}
Note that the function $f(\hat{\pi}^2)$ is not squared, since in Eq. \eqref{bige} the expectation value is taken on single spin operators rather than on the product of two of them as in Eq. \eqref{tsir}. Clearly, the effect predicted by the MLQM-induced modification of the spin is too tiny to be detected within the current experimental limits. 


Yet, the experimental framework that we have described above allows to extract meaningful information, as one can in principle infer an upper bound for the deformation parameter appearing in Eqs. \eqref{doubla} and \eqref{stringa}. In particular, for neutrons of mass $M\sim \mathcal{O}(1)$ GeV and kinetic energy $E_{\mathrm{kin}}\sim\mathcal{O}(10^{-2})$ GeV, the upper bounds for $\beta_\ell$ and $\beta_q$ are
\begin{equation}\label{bounds}
\beta_\ell\lesssim10^{19}\,, \qquad \beta_q\lesssim10^{40} \, .
\end{equation}
This result yields one of the finest set of bounds available for non-gravitational experiments with non-elementary particles \cite{bounds,bounds2}.

\section{Conclusions}

In this work, we have discussed how the existence of a minimal scale of length predicted by models of quantum gravity implies a momentum-dependent modification of the internal degrees of freedom, and thus a minimal-length induced modification of non-relativistic quantum mechanics.
We have explored the consequences of this quantum-gravity induced deformation of ordinary quantum mechanics on Bell's theorem and quantum nonlocality.
We have found that the Tsirelson bound on the violation of the CHSH inequalities that holds in ordinary quantum mechanics is exceeded in minimal-length quantum mechanics, thus leading to stronger forms of quantum nonlocality. In addition, we have investigated an experimental setting based on neutron interferometry in order to probe the tiny effects implied by the existence of a minimal length on ordinary quantum processes in non-relativistic regimes. 


In the present context, we emphasize that the modification of the canonical commutation relation between the position and momentum operators does not contradict any fundamental axiom of quantum theory. Therefore, our main result as expressed by Eq. \eqref{tsir} represents a generalization rather than a violation of the Tsirelson bound because it provides a prescription to account for the true physical observables in the presence of quantum gravitational effects.

Our results seem to suggest that the existence of a minimal length might be at odds with one or more of the principles that justify the very existence of the Tsirelson bound in quantum mechanics \cite{axiom1,axiom2,axiom3}. This conundrum has no apparently obvious resolution:
either the axioms of no-advantage for nonlocal computation postulate \cite{axiom1}, information causality \cite{axiom2} and macroscopic locality \cite{axiom3} are amenable to further generalizations, or they might be intimately related to the structure of ordinary quantum mechanics with no minimal length. In the latter instance, the maximum degree of nonlocality allowed by standard quantum mechanics, as quantified by the Tsirelson bound, would be only a lowest-order approximation to stronger forms of nonlocality featured by more general theories, corresponding perhaps to non-separable Hilbert spaces or even more exotic structures of physical states. 

Further in-depth investigations will be needed in order to clarify the basic tenets of nonlocality in physics and in particular the relation between quantum nonlocality and quantum gravity phenomenology encompassing the existence of a minimal scale of length.

\subsection*{Acknowledgements}
F.I. and L.P. acknowledge support by MUR (Ministero dell'Universit\`a e della Ricerca) via the project PRIN 2017 ``Taming complexity via QUantum Strategies: a Hybrid Integrated Photonic approach'' (QUSHIP) Id. 2017SRNBRK. F.W. is thankful to the quantum gravity group at CP$^3$ origins, University of Southern Denmark, for its kind hospitality during his research visit, and was supported by the Polish National Research and Development Center (NCBR) project ''UNIWERSYTET 2.0. --  STREFA KARIERY'', POWR.03.05.00-00-Z064/17-00 (2018-2022). L.P. is grateful to the ``Angelo Della Riccia'' foundation for the awarded fellowship received to support the study at Universit\"at Ulm. P.B., F.W. and L.P. acknowledge networking support by the COST Action CA18108.

\subsection*{Data Availability}
Data sharing not applicable to this article as no datasets were generated or analyzed during the current study.

\subsection*{Author Contributions}
All authors contributed to the original idea and to the first manuscript draft. P.B., L.P., and F.W. performed calculations; F.I. supervised the work and revised the manuscript. All authors discussed the results and edited the paper.

\subsection*{Competing interests}
The authors declare no competing interests.


\providecommand{\href}[2]{#2}\begingroup\raggedright\endgroup
\end{document}